\documentstyle[preprint,floats,pra,aps,epsfig,psfig]{revtex}

\newcommand{\bce}{\begin{center}}
\newcommand{\ece}{\end{center}}
\newcommand{\be}{\begin{equation}}
\newcommand{\ee}{\end{equation}}
\newcommand{\bea}{\vspace{0.25cm}\begin{eqnarray}}
\newcommand{\eea}{\end{eqnarray}}

\def\PRA{{Phys. Rev.} A }

\begin{document}
\centerline{ \large \bf First experimental test of Bell inequalities performed} \centerline{\large \bf using a non-maximally entangled state }

\vskip 2cm

\centerline{ M.Genovese \footnote{ genovese@ien.it. Tel. 39 011 3919234, fax 39 011 3919259},G.Brida, C.Novero }

\centerline{ Istituto Elettrotecnico Nazionale Galileo Ferraris, Str. delle Cacce\\
91,I-10135 Torino }

\centerline{E. Predazzi}

\centerline{ Dip. Fisica Teorica Univ. Torino e INFN, via P. Giuria 1, I-10125 Torino }

\vskip 2cm

\centerline{ \large Abstract}

We describe the realisation of a new test of Bell inequalities using a new scheme obtained by the superposition of type I parametric down conversion produced in two different
non-linear crystals pumped by the same laser, but with different
polarisations. This experiment is the first test of Bell inequalities using a non-maximally entangled state and thus represents an important step in the direction of  eliminating  the detection loophole.

\newpage

The idea that Quantum Mechanics (QM) could be an incomplete theory, representing a statistical approximation of a complete deterministic theory (where observable values are fixed by some hidden variable) appeared already in 1935 thank to the celebrate Einstein-Podolsky-Rosen paper \cite{EPR}. 
 
Historically, the quest for hidden variable theories  stopped
when Von Neumann published a theorem asserting the
impossibility of constructing a hidden variable theory reproducing all the
results of QM. For long time, the prestige of Von Neumann
led to an acritical acceptation of this theorem, but it was then
discovered that one of his hypotheses was too restrictive so that the
program of a hidden variable theory was still possible.

The next fundamental progress in discussing possible extensions of QM was the
discovery of Bell \cite{bell} that any realistic Local Hidden Variable LHV theory must
satisfy certain inequalities which can be violated in QM leading in principle to a possible  experimental test of the validity of QM as compared to LHV.

Since then, many interesting experiments have been devoted to a test of Bell inequalities \CITE{Mandel,asp,franson,type1,type2}, leading to a substantial agreement with quantum mechanics and disfavouring LHV theories, but, so far, no experiment has yet been able to exclude definitively such theories. In fact, so far,  one has always been forced to introduce a further additional hypothesis \CITE{santos}, due to the low total detection efficiency, stating that the observed sample of particle pairs is a faithful subsample of the whole. This problem is known as { \it detection or efficiency loophole}. The research for new experimental configurations able to overcome the detection loophole is of course of the greatest interest. 

In the 90's big progresses in this direction have been obtained by using parametric down conversion (PDC) processes. 

This technique \CITE{Mandel} has been largely employed for generating  "entangled" photon pairs, i.e. pairs of photons described by a common wave function which cannot be factorized  into the product of two distinct wave functions pertaining to separated photons. 

The generation of entangled states by parametric down conversion (PDC) has replaced other techniques, such as the radiative decay of excited atomic states, as it was in the celebrated experiment of A. Aspect et al. \CITE{asp}, for it overcomes some former limitations. In particular, having angular correlations better than 1 mrad, it overcomes the poor angular correlation of atomic cascade photons, that was at the origin of the small total efficiency of this type of experiments in which one is forced to select a small subsample of the produced photons, leading inevitably to  the detection loophole. 

The first experiments using this technique, where performed with type I PDC, which gives phase and momentum entanglement and can be used for a test of Bell inequalities using two spatially separated interferometers \cite{franson}, as realised by \cite{type1}. The use of beam splitters, however, strongly reduces the total quantum efficiency. 

In alternative, one can generate a polarisation entangled state \cite{ou}. It appears, however, that the creation of couples of photons entangled from the point of view of polarisation, which is by far the most diffuse case due to the easy experimental implementation, still suffers severe limitations, as it was pointed out recently in the literature \cite{santos}. The essence of the problem is that in generating this state, half of the initial photon flux is lost (in most of the used configurations), and one is, of necessity, led to assume that the photon's population actually involved in the experiment is a faithful sample of the original one, without eliminating the efficiency loophole. 

Recently, an experiment where a polarisation entangled state is directly generated, has been realised using Type II PDC \cite{type2}. This scheme has permitted, at the price of delicate compensations for having identical arrival time of the ordinary and extraordinary photon, a much higher total efficiency than the previous ones, which is, however, still far from the required value of $0.81$. Also, some recent experiments studying equalities among correlations functions rather than Bell inequalities \cite{dem} are far from solving these problems \cite{garuccio}. A large interest remains therefore for new experiments increasing total quantum efficiency in order to reduce and finally overcome the efficiency loophole. 

Some years ago, a very important theoretical step in this direction has been performed recognising that, while for maximally entangled pairs a total efficiency larger than to 0.81 is required to obtain an efficiency-loophole free experiment, for non maximally entangled pairs this limit is reduced to 0.67 \cite{eb} (in the case of no background). However, it must be noticed that, for non-maximally entangled states, the largest discrepancy between quantum mechanics and local hidden variable theories is reduced: thus a compromise between a lower total efficiency and a still sufficiently large value of this difference will be necessary when realising of an experiment addressed to overcome the detection loophole. 

Considering a polarization entangled state of photons of the form

\begin{equation}
\vert \psi \rangle = {\frac{ \vert H \rangle \vert H \rangle + f \vert V
\rangle \vert V \rangle }{\sqrt {(1 + |f|^2)}}}  \label{eq:Psi}
\end{equation}
where $H$ and $V$ indicate horizontal and vertical polarisations respectively, the parameter $f$ describes how much the state \ref{eq:Psi} differs from a maximally entangled one. 

The region corresponding to the one where the detection loophole is eliminated in the plane $f$ and $\eta$ (total detection efficiency) is shown in fig.1.

Let us now describe  more in detail our experimental set-up. It derives from developing 
\cite{napoli} a proposal  made by Hardy \cite
{hardy} and  is based on the creation of a polarisation (non maximally-)
entangled states of the form \ref{eq:Psi}
via the superposition of the spontaneous fluorescence emitted
by two non-linear crystals driven by the same pumping laser. The
crystals are put in cascade along the propagation direction of the pumping
laser and the superposition is obtained by using an appropriate optics.

More in details (see fig. 2),  two crystals of $LiIO_3$ (10x10x10 mm) are placed along the pump laser propagation, 250 mm apart,  a distance smaller than the coherence length of the pumping laser. This guarantees indistinguishibility in the creation of a couple of photons in the first or in the second crystal. A couple of planoconvex lenses of 120 mm focal length centred in between, focalises the spontaneous emission from the first crystal into the second one maintaining exactly, in principle, the angular spread. A hole of 4 mm diameter is drilled into the centre of the lenses in order to allow transmission of the pump radiation without absorption and, even more important, without adding stray-light, because of fluorescence and diffusion of the UV radiation.   This configuration, which realises  what is known as  an "optical condenser", was chosen among others, using an optical simulation program, as a compromise between minimisation of aberrations (mainly spherical and chromatic) and losses due to the number of optical components. The pumping beam at the exit of the first crystal is displaced from its input direction by birefringence: the small quartz plate (5 x5 x5 mm) in front of the first lens of the condensers compensates this displacement, so that the input conditions are prepared to be the same for the two crystals, apart from alignment errors. Finally, a half-wavelength plate immediately after the condenser rotates the polarisation of the Argon beam and excites in the second crystal a spontaneous emission cross-polarised with respect to the first one. With a phase matching angle of $51^o$, the spontaneous emissions at 633 and 789 nm (which are the wave lengths to be used for the test) are emitted at $3.5^o$ and $4^o$ respectively. The dimensions and positions of both plates are carefully chosen in order that they do not  intersect this two conjugated emissions. 

We have used as photo-detectors two avalanche photodiodes with active quenching (EG\&G SPCM-AQ) with a sensitive area of 0.025 $mm^2$ and dark count below 50 counts/s.
The PDC signal was coupled to an optical fiber (carrying the light to  the detectors) by means of a microscope objective with magnification 20.
The quantum efficiency, included the fiber coupling, has been measured to be $0.535 \pm 0.008$ at 633 nm  \cite{brida}.

The output signals from the detectors are pulses of Transistor-Transistor-Logic (TTL) like amplitude levels which are routed  to a two channel counter, in order to have the number of events on single channel, and to a  
Time to Amplitude Converter (TAC) circuit, followed by a single channel analyzer, for selecting and counting coincidence events. 

A very interesting degree of freedom of this configuration is given by the fact that by tuning the pump intensity between the two crystals, one can easily
tune the value of $f$, which determines how far from a maximally
entangled state ($f=1$) the produced state is. This is a fundamental property,
which permits to select the most appropriate state for the experiment.

The main problem of this configuration is the alignment, which is of the utmost relevance for a high visibility. 
The solution of this problem lies in  a technique, that  had been already applied  in our laboratory \cite{brida} for metrological studies, namely the use of  an optical amplifier
scheme, where a solid state laser is injected into the crystals together with
the pumping laser, an argon laser at 351 nm wavelength (see fig.1). If the angle of injection is selected appropriately, a stimulated emission along the correlated direction appears, allowing an easy identification of the two correlated directions. Then, stopping the stimulated emission of the first crystal, and rotating the polarisation of the diode laser one obtaines the stimulated emission in the second crystal and can check the superposition with the former.

We think that the proposed scheme will lead to a further step towards a conclusive experimental test of non-locality in quantum mechanics. The
analysis of the experiments realised up to now \cite{santos} shows in fact
that visibility of the wanted effect (essentially visibility of interference
fringes) and overall quantum detection efficiency are the main parameters
in such experiments. One first advantage of the proposed configuration with
respect to most of the previous experimental set-ups is that all the
entangled pairs are selected (and not just $< 50 \%$ as with beams splitters); furthermore, it does not require delicate compensations for the optical paths of the ordinary and extraordinary rays emerging from the crystal.

For this time being, the results which we are going to present are still far from a definite solution of the detection loophole; nevertheless, being the first test of Bell inequalities using a non-maximally
entangled state, they
represents an important step in this direction. Furthermore, this
configuration allows to use any pair of correlated frequencies and not only
the degenerate ones. We have thus realised this test using for a first time
two different wave lengths (at $633$ and $789$ nm).

An experiment which presents analogies with our, has been realised recently in ref. \cite{Kwiat}.
The main difference between the two experiments is that in \cite{Kwiat} the
two crystals are very thin and in contact with orthogonal optical
axes: this allows a "partial" superposition of the two emissions with
opposite polarisation. This overlapping is mainly due to the finite dimension of the pump laser beam, which reflects into a finite width of each wavelength emission.
A much better superposition can be obtained with the present scheme, by fine tuning the crystals' and optics' positions and using the parametric amplifier trick. 
  Furthermore, in the experiment of ref. \cite{Kwiat}
the value of $f$ is in principle tunable by rotating the
polarisation of the pump laser; however, this reduces the power of the
pump producing PDC already in the first crystal, while in our case the whole pump
power can always be used in the first crystal, tuning the PDC produced in
the second one.

As a first check of our apparatus, we have measured the interference
fringes, varying the setting of one of the polarisers, while leaving the other
fixed. We have found a high visibility, $V=0.973 \pm 0.038$.

Our results are summarised by the value obtained for the Clauser-Horne sum,

\begin{equation}
CH=N(\theta _{1},\theta _{2})-N(\theta _{1},\theta _{2}^{\prime })+N(\theta
_{1}^{\prime },\theta _{2})+N(\theta _{1}^{\prime },\theta _{2}^{\prime
})-N(\theta _{1}^{\prime },\infty )-N(\infty ,\theta _{2})  \label{eq:CH}
\end{equation}
which is strictly negative for local realistic theory. In (\ref{eq:CH}), $N(\theta _{1},\theta _{2})$ is the number of coincidences between
channels 1 and 2 when the two polarisers are rotated to an angle $\theta _{1}$
and $\theta _{2}$ respectively ($\infty $ denotes the absence of selection
of polarisation for that channel)

On the other hand,  quantum mechanics predictions for $CH$ can be larger than zero: for a maximally
entangled state the largest value is obtained for $\theta _{1}=67^{o}.5$ , $%
\theta _{2}=45^{o}$, $\theta _{1}^{\prime }=22^{o}.5$ , $\theta _{2}^{\prime
}=0^{o}$ and corresponds to a ratio 
\begin{equation}
R=[N(\theta _{1},\theta _{2})-N(\theta _{1},\theta _{2}^{\prime })+N(\theta
_{1}^{\prime },\theta _{2})+N(\theta _{1}^{\prime },\theta _{2}^{\prime
})]/[N(\theta _{1}^{\prime },\infty )+N(\infty ,\theta _{2})]  \label{eq:R}
\end{equation}
equal to 1.207.

For non-maximally entangled states the angles for which CH is maximal are
somehow different and the maximum is reduced to a smaller value. The angles corresponding to
the maximum can be evaluated maximising Eq. \ref{eq:CH} with 

\bea
\left. \begin{array}{l}

  N[\theta _{1},\theta _{2}] =  [ \epsilon _1^{||} \epsilon _2^{||} (Sin[\theta _{1}]^{2}\cdot Sin[\theta_{2}]^{2}) + \\
  \epsilon _1^{\perp} \epsilon _2^{\perp} 
(Cos[\theta _{1}]^{2} \cdot Cos[\theta _{2}]^{2} )\\
  (\epsilon _1^{\perp} \epsilon _2^{||} Sin[\theta _{1}]^2\cdot Cos[\theta _{2}]^2 + \epsilon _1^{||} \epsilon _2^{\perp} 
Cos[\theta _{1}]^2 \cdot Sin [\theta _{2}]^2 )  \\
 + |f|^{2}\ast (\epsilon _1^{\perp} \epsilon _2^{\perp} (Sin[\theta _{1}]^{2}\cdot Sin[\theta_{2}]^{2}) +  \epsilon _1^{||} \epsilon _2^{||}
(Cos[\theta _{1}]^{2} Cos[\theta _{2}]^{2} ) +\\
(\epsilon _1^{||} \epsilon _2^{\perp} Sin[\theta _{1}]^2\cdot Cos[\theta _{2}]^{2} +\\
 \epsilon _1^{\perp} \epsilon _2^{||} 
Cos[\theta _{1}]^2 \cdot Sin [\theta _{2}]^2 )   \\
  +  (f+f^{\ast }) ((\epsilon _1^{||} \epsilon _2^{||} + \epsilon _1^{\perp} \epsilon _2^{\perp} - \epsilon _1^{||} \epsilon _2^{\perp} - 
\epsilon _1^{\perp} \epsilon _2^{||})
\cdot (Sin[\theta _{1}]\cdot Sin[\theta _{2}]\cdot
Cos[\theta _{1}]\cdot Cos[\theta _{2}]) ]  /(1+|f|^{2}) \,
\end{array}\right. \, .   
\eea
where (for the case of non-ideal polariser) $\epsilon _i^{||}$ and $\epsilon _i^{\perp}$ 
correspond to the transmission when the polariser (on the branch $i$)  axis is aligned or normal to the polarisation axis respectively.

In our case, we have generated a state with $f \simeq 0.4$ which corresponds,
for $\theta_1 =72^o.24$ , $\theta_2=45^o$, $\theta_1 ^{\prime}= 17^o.76$ and 
$\theta_2 ^{\prime}= 0^o$, to $R=1.16$.

Our experimental result is $CH = 512 \pm 135$ coincidences per second, 
which is almost four standard deviations different from zero and compatible with the theoretical value predicted by quantum mechanics.
In terms of the ratio (\ref{eq:R}), our result is $1.082 \pm 0.031$.

For the sake of comparison, one can consider the value obtained with the angles which optimise Bell inequalities violation for a maximally entangled state (given before Eq. \ref{eq:R}).
The result is $CH = 92 \pm 89$, which, as expected, shows a smaller violation than the value obtained with the correct angles setting.

In summary, this is the first measurement of the violation of the Clauser-Horne inequality (or other Bell inequalities) using a non-maximally entangled state and thus represents and interesting result as a first step in the direction of eliminating the detection
loophole. Further developments in this sense are the purpose of this collaboration.

\bigskip 

\vskip1cm \noindent {\bf Acknowledgements} \vskip0.3cm 
We would like to
acknowledge the support of the Italian Space Agency under contract LONO 500172 and
of  MURST via special programs "giovani ricercatori", Dip. Fisica Teorica Univ. Torino.

\vfill \eject

\end{document}